\documentclass[9.5pt,journal,final,letterpaper,twocolumn]{IEEEtran}

\usepackage{amsmath,amssymb}
\usepackage{graphicx,verbatim,setspace}

\newtheorem{thm}{Theorem}
\newtheorem{lem}[thm]{Lemma}

\newtheorem{cor}[thm]{Corollary}

\newtheorem{rem}{Remark}
\newtheorem{example}{Example}

\newcommand\diag{\operatorname{diag}}

\newcommand\rank{\operatorname{rank}}

\newcommand\Cov{\operatorname{Cov}}
\newcommand\eCov{\operatorname{eCov}}
\newcommand\sCov{\operatorname{sCov}}

\begin{document}

\title{The Identifiability of Covarion Models in Phylogenetics}

\author{Elizabeth S.~Allman, John A.~Rhodes
\thanks{Department of Mathematics and Statistics, University of
Alaska Fairbanks, PO Box 756660, Fairbanks, AK 99775-6660;
e.allman@uaf.edu, j.rhodes@uaf.edu}
\thanks{The authors thank the Isaac Newton Institute and
the National Science Foundation. Parts of this work were conducted
during residencies at INI, and with support from NSF grant DMS
0714830.} }

\maketitle

\begin{abstract}
Covarion models of character evolution describe inhomogeneities in
substitution processes through time. In phylogenetics, such models
are used to describe changing functional constraints or selection
regimes during the evolution of biological sequences. In this work
the identifiability of such models for generic parameters on a known
phylogenetic tree is established, provided the number of covarion
classes does not exceed the size of the observable state space.
`Generic parameters' as used here means all parameters except possibly those
in a set of measure zero within the parameter space.
Combined with earlier results, this implies both the tree and
generic numerical parameters are identifiable if the number of
classes is strictly smaller than the number of observable states.
\end{abstract}

\begin{keywords}
phylogenetics, Markov processes on trees, covarion models,
statistical consistency
\end{keywords}
\section{Introduction}

Phylogenetic inference is now generally performed in a statistical
framework, using probabilistic models of the evolution of biological
sequences, such as DNA or proteins. To rigorously establish the
validity of such an approach, a fundamental question that must be
addressed is whether the models in use are \emph{identifiable}: From
the theoretical distribution predicted by the model, is it possible
to uniquely determine all parameters? Parameters for simple models
include the topology of the evolutionary tree, edge lengths on the
tree, and rates of various types of substitution, though more
complicated models have additional parameters as well. If a model is
non-identifiable, one cannot show that performing inference with it
will be \emph{statistically consistent}. Informally, even with large
amounts of data produced by an evolutionary process that was
accurately described by the model, we might make erroneous
inferences if we use a non-identifiable model.

Identifiability for the most basic phylogenetic models, such as the
Jukes-Cantor, Kimura, and all other time-reversible models, follows
from Chang's work on the general Markov model \cite{MR97k:92011}.
However, for models with rate variation across sites, where the
distribution of rates is not fully known, only recently have the
first positive results been obtained
\cite{ARidtree,ARGMI,AllmanAneRhodes08}. Despite its widespread use
in data analysis, identifiability of the GTR+$\Gamma$+I model has
yet to be addressed rigorously. (Unfortunately the proof of
identifiability given in \cite{Rog01} has fundamental gaps, as
explained in the appendix of \cite{AllmanAneRhodes08}.)

\smallskip

The covarion model, introduced in its basic mathematical form by
Tuffley and Steel \cite{MR1604518}, incorporates rate variation
within lineages rather than across sites. Extensions of the basic
version of the model have appeared in a variety of analyses of
experimental data, with authors referring to the model using
terminology such as `covarion' \cite{HuelCov}, `covarion-like'
\cite{Galt,Wang07}, `site-specific rate variation' \cite{Galt,GRDH},
`Markov-modulated Markov process' \cite{GaltJM,GasGui}, or `temporal
hidden Markov models' \cite{Whelan}. We use the name `covarion' in
this paper for simplicity, although we acknowledge the model does
not capture the full complexity of the process originally proposed
by Fitch and Markowitz \cite{FM}. Informally, the covarion model
allows several classes (\textit{e.g.}, invariable, slow, and fast),
with characters evolving so they not only change between observable
states, but also between classes. Though the class is never
observed, it affects the evolutionary process over time. The model
thus attempts to capture the fact that substitution rates may speed
up or slow down at different sites in a sequence at different times
in their descent. Changing functional constraints or selection
regimes are possible sources of such a process.

Identifiability of even the tree parameter under the covarion model
was not established with its introduction in \cite{MR1604518},
despite strong efforts.  In \cite{ARidtree}, the authors established
that for generic choices of covarion parameters tree topologies are
indeed identifiable, provided the number of covarion classes is less
than the number of observable states. Thus for nucleotide models of
DNA there can be 3 classes, though for amino acid models of proteins
one can allow 19 classes, and for codon models of DNA up to 60
classes. `Generic' here means that there could be some parameter
choices for which identifiability fails, though they will be rare
(of Lebesgue measure zero). In fact, if parameters are chosen
randomly, with any natural notion of random, one can be sure the
tree topology is identifiable.

\smallskip

Since the notion of generic identifiability is perhaps not widely known, and will play a key role in this work as well, we elaborate on its meaning. For statistical models in general, it is most desirable to establish identifiability over the full parameter space. However, such a strong claim may not hold, so that the best possible result is to establish identifiability over most of the parameter space, and completely characterize all those parameter choices for which identifiability fails. Generic identifiability results are a little weaker than this, in that while identifiability is established over most of the parameter space, they allow for ignorance about identifiability on a small subset of the parameter space. This exceptional subset of parameter space contains all parameters for which identifiability fails, but may also contain some parameters that are identifiable. Complex statistical models can be quite difficult to analyze, so that generic identifiability is sometimes the strongest known result. For instance, though hidden Markov models are widely used in bioinformatics and other fields, and generic identifiability was proved for HHMs in \cite{Petrie}, we know of no improvements on that work in the nearly 40 years since it appeared. Phylogenetic models are similar to HMMs in that they posit unobserved variables, at the internal nodes of a tree, but typically have more complex parameterizations than HMMs. Thus we consider their
analysis to be even more challenging.

\smallskip

The question of identifiability of numerical parameters for the
covarion model was left open by \cite{ARidtree}. In this article, we assume the
tree topology is known, and establish identifiability of the
numerical parameters of several variants of the covarion model for
generic parameter choices, provided the number of covarion classes
is strictly less than the number of observable states. For certain
versions of a covarion model, this can be strengthened to allow
one more class, so that the number of classes and observable states
may be the same.

We consider three variants of the covarion model, which extend the
Tuffley-Steel model, and have previously appeared in works of
others, though without our formal terminology: The \emph{scaled
covarion} model, $\sCov$, assumes all classes undergo substitutions
according to a common process but at rescaled rates. The \emph{equal
stationary distribution covarion model}, $\eCov$, generalizes this
to allow in-class substitution processes to vary more across
classes, provided they have identical stationary distributions and
class change rates are independent of the base. Finally, in the
\emph{general covarion model}, $\Cov$, each class may undergo
substitutions quite differently as long as the entire process is
time reversible. $\Cov$ is the model described in \cite{Whelan},
$\eCov$ is developed in \cite{GasGui}, and $\sCov$ is used in
\cite{GRDH}.

Note these models are nested, $$\sCov\subset\eCov\subset \Cov,$$
though each submodel is non-generic within its supermodels. Because
identifiability is established here only for generic parameters, it
is necessary to state and prove the generic identifiability of all
three covarion models to encompass the range of models used in
practice.

\smallskip

In Section \ref{sec:param} we formally present these models, and in
Section \ref{sec:results} we state our results precisely. That
section also provides an overview of the proof. For those whose
primary interest is understanding the result, and who do not wish to
delve into the full mathematical arguments behind it, we suggest
that reading through Section \ref{sec:results} may suffice. The
remainder of the paper provides the rather detailed arguments that
are essential to rigorously establishing identifiability.

We also note that many practitioners have conducted data analysis
with models combining covarion features with across-site
rate variation, such as that modeled by a discrete $\Gamma$
distribution. While the identifiability of such models has not been
established rigorously as of yet, we view the main theorems of this
paper as providing a first step toward understanding of these more
complex models.

\smallskip

This work was influenced by many useful discussions concerning
covarion models that we had with participants of the Isaac Newton
Institute's Programme in Phylogenetics. Simon Whelan deserves
particular thanks for explaining his forthcoming work \cite{Whelan}.

We also thank the referees for their helpful suggestions, and especially
Christopher Tuffley, who
noted a flaw in an earlier version of Section \ref{sec:covgen}, and suggested
the simpler argument that appears there now.

\section{The Parameterization of the Covarion
Models}\label{sec:param}

For the purpose of orientation,
we briefly recall a simpler phylogenetic model, the $\kappa$-state general time reversible (GTR) model. The basic state change process is specified by a $\kappa\times\kappa$
rate matrix $Q$, whose off-diagonal $i$,$j$-entry gives an instantaneous rate $(>0)$ at which a character in state $i$ enters state $j$. Each row of $Q$ must add to 0. As a consequence, $Q$ has a unique left eigenvector $\boldsymbol \pi$ with eigenvalue 0, the stationary vector for $Q$.  Time reversibility is mathematically formulated as the assumption that $\diag(\boldsymbol \pi)Q$ is symmetric. Character change along a rooted metric tree $T$ is then modeled as follows: The entries of $\boldsymbol \pi$ give the probability that a character is in the various states at the root of the tree. Along each edge $e$ of $T$, directed away from the root, the conditional probabilities of state changes are given by the Markov matrix $M_e=\exp(Qt_e)$, where $t_e\ge0$ is the edge length. From this information one can compute
the probability of any specification of states at the leaves of the tree. Due to the time reversibility assumption, the location of the root within the tree actually has no effect on this probability distribution. Thus the parameters of the model are the topology of the unrooted tree $T$, the collection of edge lengths $\{t_e\}$, and the rate matrix $Q$.

To present the covarion models, we first focus on the process of state change.
It will be convenient to adopt
terminology most appropriate to nucleotide sequences. In particular, in discussing covarion models
we limit our use of the word `state' which is commonly used for all
Markov models, because the number of states at internal nodes of a tree
differs from that at leaves, even though there is a relationship
between them. We instead refer to observable states as `bases,' and
to rate classes as `classes.' Thus at a leaf a state is simply a
base, while at an internal node a state is a pair of a class and a
base. We caution the reader that this usage of `base' is not
standard in biology, as it encompasses the 4 bases in nucleotide
sequences, as well as the 20 amino acids of protein sequences, and
the 61 codons in a model of codon substitution. Also, while it is
often natural to think of `classes' as being associated to rate
scalings, this may be misleading, as several of the models we
formalize allow for more generality. We use $[\kappa]=\{1,2,\dots,\kappa\}$
to denote the set of bases
and $[c]=\{1,2,\dots, c\}$ to denote the set of classes.

\smallskip

To refer to entries of vectors and matrices of size $c\kappa$, it
will be convenient to index entries using interchangeably the set
$[c\kappa]$, and the set $[c]\times[\kappa]$
with lexicographic order. Thus the index $(i,j)$, which should be
interpreted as the `class $i$, base $j$' index, is equivalent to
$(i-1)c+j$. Entries in a $c\kappa\times c\kappa$ matrix, then, can
be referred to by an ordered pair of indices, each of which is an
ordered pair in $[c]\times[\kappa]$.

\smallskip

Let $c,\kappa$ be positive integers. The most general $c$-class,
$\kappa$-base covarion model, introduced by Whelan in \cite{Whelan},
is specified in the following way:

\begin{enumerate}

\item[(1)] For each $i\in[c]$, a base-change process
for class $i$ is described by a rate-matrix $Q_i$ of size $\kappa
\times \kappa$. We assume all $Q_i$ are distinct, so that no two
classes undergo substitutions at the same rates. For $c-1$ values of
$i$ we require that the off-diagonal entries of $Q_i$ are strictly
positive so that all substitutions are possible, and the rows sum to
0. For the remaining $Q_i$ we only require that all off-diagonal
entries be non-negative and that rows sum to 0. In particular, we
allow $Q_i$ for at most one $i$ to be the 0-matrix, in order to
model an invariable class.

\item[(2)] For each ordered pair of classes $i_1\ne i_2$, a diagonal matrix
$S_{i_1i_2}$ of size $\kappa\times\kappa$ describes switching rates
from class $i_1$ to class $i_2$. The entries of $S_{i_1i_2}$ are
non-negative. The requirement that $S_{i_1i_2}$ be diagonal will imply
that instantaneous base switches
do not occur simultaneously with class switches.

\item[(3)] Let $R$ be the $c\kappa\times c\kappa$ matrix which, when
viewed in $c\times c$ block form, has as its off-diagonal
$i_1,i_2$-block $S_{i_1i_2}$ and as its $i$th diagonal block
$Q_i-\sum_{i_2} S_{ii_2}$. Note each row of $R$ sums to 0. We
require that $R$ describe a time-reversible process; that is, for
some vector $\boldsymbol \mu$ with positive entries summing to 1 the
matrix
$$\diag(\boldsymbol \mu) R$$
is symmetric.

\end{enumerate}

We may rescale $R$, or equivalently all entries of the $Q_i$ and
$S_{i_1i_2}$, so that $$\operatorname{trace}(\diag(\boldsymbol
\mu)R)=-1.$$ Requiring this normalization avoids a trivial non-identifiability
issue in which rescaling of edge lengths would have the same effect as rescaling $R$.
It also imposes a scale on edge lengths so that the average instantaneous
rate of (base,class) changes under the Markov process is 1 per unit of edge length.
We will assume throughout the rest of this paper that
this normalization has been made. Consequently, if two such
matrices are multiples of one another, we may conclude they are
equal.

Any matrix $R$ with these properties will be called a \emph{covarion
rate matrix} for the \emph{general covarion model},
$\Cov(c,\kappa)$, with $c$ classes and $\kappa$ bases.

\smallskip

We may write $$\boldsymbol \mu=(\sigma_1 \boldsymbol \pi_1, \,
\sigma_2 \boldsymbol \pi_2,\, \dots,\, \sigma_c \boldsymbol \pi_c)$$
where the $\boldsymbol \pi_i\in \mathbb R^\kappa$ and $\boldsymbol
\sigma=(\sigma_1,\dots,\sigma_c)\in \mathbb R^c$ are vectors of
positive entries summing to 1. Then the symmetry of
$\diag(\boldsymbol \mu)R$ implies the symmetry of $\diag(\boldsymbol
\pi_i)Q_i$ for each $i$. Thus our assumptions ensure the $Q_i$ each
define time-reversible processes. Additionally we find
\begin{equation}
\sigma_{i_1} \diag(\boldsymbol \pi_{i_1})S_{i_1i_2}= \sigma_{i_2}
\diag(\boldsymbol \pi_{i_2})S_{i_2i_1}.\label{eq:switch}
\end{equation}
These conditions are equivalent to the time-reversibility of
$R$.

\smallskip
A specialization of $\Cov(c,\kappa)$ described in \cite{GasGui}
assumes further that
\begin{enumerate}

\item[(4)] The base substitution processes described by the $Q_i$
have equal stationary distributions, $\boldsymbol \pi_i=\boldsymbol
\pi$.

\item[(5)] The switching matrices $S_{i_1i_2}$ are scalar, so
$S_{i_1i_2}=s_{i_1i_2} I_\kappa$, where $I_\kappa$ is the
$\kappa\times\kappa$ identity matrix.
\end{enumerate}
We refer to this as the \emph{equal stationary distribution covarion
model}, denoted by $\eCov(c,\kappa)$.

The model $\eCov(c,\kappa)$ can also be conveniently described in
tensor notation. For any vectors or matrices $A=(a_{i_1i_2})$ and
$B=(b_{j_1j_2})$, let $A\otimes B$ denote the tensor, or Kronecker,
product. Using ordered-pair indices as above, we order rows and
columns of $A\otimes B$ so the $(i_1,j_1),(i_2,j_2)$ entry is
$a_{i_1i_2}b_{j_1j_2}$. With the class switching process for $\eCov$
specified by a $c\times c$ rate matrix $S$ with off-diagonal entries
$s_{i_1i_2}$, and rows summing to 0, then \begin{gather*} R=
\diag(Q_1,Q_2,\dots, Q_c)+S\otimes I_\kappa,\\
\boldsymbol\mu=\boldsymbol \sigma\otimes\boldsymbol \pi.
\end{gather*}
The symmetry of $\diag(\boldsymbol \mu) R$ is equivalent to the
symmetry of each $\diag(\boldsymbol \pi)Q_i$ and of
$\diag(\boldsymbol \sigma) S$. Thus the class switching process
described by $S$ is time-reversible as well.

\medskip

A further specialization from $\eCov$ yields the \emph{scaled
covarion model}, $\sCov(c,\kappa)$, which assumes
\begin{enumerate}
\item[(6)] For some rate matrix $Q$ and distinct non-negative $r_1,r_2,\dots, r_c$, $Q_i=r_iQ.$
\end{enumerate}
For this submodel, the full covarion process has rate matrix
\begin{gather}
R= \diag(r_1,r_2,\dots, r_c)\otimes Q+S\otimes I_\kappa.\label{eq:RsCov}
\end{gather}

\begin{example}
$\sCov(2,4)$ is just a generalization of the
Tuffley-Steel covarion model of nucleotide substitution \cite{MR1604518}. For any $s_1,s_2>0$, let $$S =\begin{pmatrix} -s_1 & s_1\\
s_2&-s_2\end{pmatrix},\ \ \ \boldsymbol
\sigma=(\sigma_1,\sigma_2)=\left ( \frac{s_2}{s_1+s_2},
\frac{s_1}{s_1+s_2}\right ).$$ Then $S$ defines a time-reversible
switching process with stationary vector $\boldsymbol \sigma$. For
any $Q, \boldsymbol \pi$ of a 4-base GTR model, taking $1=r_1>r_2$
we obtain a rate matrix with block structure
$$
\lambda
\begin{pmatrix}
Q - s_1I & s_1I\\
s_2I & r_2Q - s_2I
\end{pmatrix},
$$ while
$$\boldsymbol \mu= ( \sigma_1 \pi_1, \ \sigma_1 \pi_2, \ \sigma_1
\pi_3, \ \sigma_1 \pi_4, \ \sigma_2 \pi_1, \ \sigma_2 \pi_2, \
\sigma_2 \pi_3, \ \sigma_2 \pi_4 ). $$ If $r_2=0$, then an
invariable class is included, and this is exactly the Tuffley-Steel
model.
\end{example}

\smallskip

\begin{example}
If $c\ge 3$, the requirement for $\eCov(c,\kappa)$ that the class
switching process described by $S$ be time-reversible implies
stronger relationships among its entries than merely requiring rows
sum to 0. If
$$
S = \begin{pmatrix}
-(s_{12} + s_{13}) & s_{12} & s_{13}\\
s_{21} & -(s_{21} + s_{23}) & s_{23}\\
s_{31} & s_{32} & -(s_{31} + s_{32})
\end{pmatrix},
$$
and $\boldsymbol \sigma$ are such that $\diag (\boldsymbol \sigma)
S$ is symmetric, then one can show (most easily by using symbolic
algebra software, such as Maple or Singular) that
$$s_{12}s_{23}s_{31}-s_{13}s_{21}s_{32}=0,$$
and
$$
\boldsymbol \sigma= \frac 1{s_{21} s_{32} + s_{12} s_{32} + s_{12}
s_{23}} ({s_{21} s_{32}},\ {s_{12} s_{32}},\ {s_{12} s_{23}}). $$

Let $Q_1,Q_2,Q_3$ denote $\kappa$-base GTR rate
matrices
with a common stationary vector $\boldsymbol \pi$. Then, up to a scaling factor, the matrix
$$\begin{pmatrix}
Q_1 - (s_{12} + s_{13})I & s_{12}I & s_{13}I\\
s_{21}I & Q_2 - (s_{21} + s_{23})I & s_{23}I\\
s_{31}I & s_{32}I & Q_3 - (s_{31} + s_{32})I \\
\end{pmatrix}$$
is a rate matrix for $\eCov(3,\kappa)$ with stationary vector
$$
\boldsymbol{\mu} = (\sigma_1 \boldsymbol{\pi} \ \ \sigma_2
\boldsymbol{\pi} \ \ \sigma_3 \boldsymbol{\pi} ).
$$
Such models are presented in \cite{GasGui}.
\end{example}

\smallskip

\begin{example} Let $Q_1,Q_2$ denote $\kappa$-base GTR rate
matrices, with stationary vectors $\boldsymbol \pi_1, \boldsymbol
\pi_2$. Let $\boldsymbol \sigma=(\sigma_1, \sigma_2)$ be any vector
of positive entries summing to 1, and $\boldsymbol s =
(s_1,s_2,\dots,s_\kappa)$ any vector of positive numbers. Then
defining
\begin{align*}
S_{12}&=\sigma_2 \diag(\boldsymbol \pi_2)\diag(\boldsymbol s),\\
S_{21}&=\sigma_1 \diag(\boldsymbol \pi_1)\diag(\boldsymbol s),
\end{align*}
ensures that equation (\ref{eq:switch}) is satisfied. For suitable
$\lambda$, the matrix $$\lambda \begin{pmatrix} Q_1-S_{12} &
S_{12}\\S_{21}&Q_2-S_{21}
\end{pmatrix}$$ is thus a rate matrix for the model
$\Cov(2,\kappa)$, and of the type described in \cite{Whelan}.

\end{example}

\smallskip

To specify any of the covarion models $\Cov(c,\kappa)$,
$\eCov(c,\kappa)$, or $\sCov(c,\kappa)$ on a topological tree $T$,
in addition to $R$ we must specify edge lengths $\{t_e\}$.
These determine Markov matrices $M_e$ for each edge $e$ of the tree as follows: For
every internal edge $e$ of the tree, $M_e=\exp(Rt_e)$ is $c\kappa\times c\kappa$
and describes
(class,\,base)-substitutions over the edge. Letting $\mathbf
1_c=\begin{pmatrix} 1&1&\dots&1
\end{pmatrix}\in\mathbb R^c$ be a row vector, and $I_\kappa$
the $\kappa\times\kappa$ identity, set
$$J=\mathbf 1_c^T\otimes I_\kappa=\begin{pmatrix}
I_\kappa&I_\kappa&\dots&I_\kappa \end{pmatrix}^T.$$ Then on every
pendant edge $e$ of the tree, $M_e=\exp(Rt_e)J$ is $c\kappa\times\kappa$. Notice that $J$ serves to hide class information,
by summing over it, so that only bases may be observed.

Because the process defined by $R$ is reversible, we may arbitrarily
choose any internal vertex of the tree as the root, and using
$\boldsymbol \mu$ as a root distribution compute the joint
distribution of bases at the leaves of the tree in the usual way for
Markovian phylogenetic models on trees. For an $n$-leaf tree, this
distribution is naturally thought of as an $n$-dimensional
$\kappa\times\kappa\times\cdots\times\kappa$ array.

\medskip

Let $P=\hat P\otimes I_\kappa $, where $\hat P$ is a $c\times c$
permutation matrix. Then replacing $R$ by $P^TRP$ simply permutes
the classes. As no information on classes is observed, it is easy to
see this has no effect on the joint distribution of bases arising
from a covarion model. Thus we must account for this trivial source
of non-identifiability. For $\sCov(c,\kappa)$ this could be done by
requiring the $r_i$ be enumerated in descending order. However, for
$\Cov(c,\kappa)$ and $\eCov(c,\kappa)$ there need not be any natural
ordering of the $Q_i$. To treat all these models uniformly, we will
seek identifiability only up to permutation of classes.

\medskip

Note that as formulated above, the covarion models generalize
mixture models on a single tree with a finite number of classes.
Indeed, one need only choose the switching matrix $S$ for $\sCov$ or
$\eCov$ to be the zero matrix, or set all $S_{i_1i_2}=0$ for $\Cov$,
to describe across-site rate variation. However, such choices are
non-generic --- of Lebesgue measure zero within the covarion models.
Since our main result allows for non-generic exceptions to identifiability,
we caution that it does not rigorously imply anything about across-site rate variation models,
though it is perhaps suggestive.

At one point in our arguments we will in fact
need an assumption that rules out consideration of across-site rate variation models.
In Lemma \ref{lem:GTRconj}, we require that the switching process for
$\Cov(c,\kappa)$ is irreducible in the following sense: Say class
$i$ \emph{communicates} to class $i'$ when all diagonal entries of
$S_{ii'}$ are positive. Then \emph{class irreducibility} of $R$ will
mean that for each pair of classes $i\ne i'$ there is a chain of
classes $i=i_0,i_1,i_2,\dots,i_n=i'$ with $i_k$ communicating to
$i_{k+1}$. For the models $\eCov$ and $\sCov$, this definition is
equivalent to the usual definition of irreducibility,
\cite{hornJohnson85}, for the Markov process described by the
switching matrix $S$. Moreover, class irreducibility of $R$,
together with the assumption that all entries of some $Q_i$ are
non-zero implies irreducibility of $R$ in the usual sense.

Note that class irreducibility holds
for generic choices of covarion parameters for all three covarion
models, as generically all diagonal entries of all $S_{ii'}$ are
non-zero. Therefore, despite its important role in establishing the results,
we do not refer to irreducibility explicitly in
statements of theorems which only make claims for generic parameter
choices.

\section{Statement of Theorems and Overview}\label{sec:results}
We establish the following:

\smallskip

\begin{thm}\label{thm:main}
Consider the models $\Cov(c,\kappa)$, $\eCov(c,\kappa)$, and
$\sCov(c,\kappa)$ on an $n$-leaf binary tree, $n\ge 7$. If the tree
topology is known, then for generic choices of parameters all
numerical parameters are identifiable, up to permutation of classes,
provided $c\le \kappa$ for $\sCov$ and $\eCov$, and provided
$c<\kappa$ for $\Cov$.
\end{thm}

\smallskip

Combined with earlier work in \cite{ARidtree}, this shows:

\smallskip

\begin{cor}\label{cor:main}
Consider the models $\Cov(c,\kappa)$, $\eCov(c,\kappa)$, and
$\sCov(c,\kappa)$ on an $n$-leaf binary tree, $n\ge 7$. Then for
generic choices of parameters, the tree topology and all numerical
parameters are identifiable, up to permutation of classes, provided
$c< \kappa$.
\end{cor}

\smallskip

In outline, the proof of the theorem is as follows: Section
\ref{sec:eig} addresses basic properties of eigenvectors and eigenvalues of a
covarion rate matrix, and discusses the
form of joint distributions from covarion models on 2-leaf trees.
This section provides preliminary results needed for the
main arguments, which span the remainder of this article.

\begin{figure}[h]
\begin{center}
\includegraphics[height=.7in]{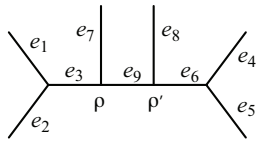}
\end{center}
\caption{The 6-leaf tree on which arguments will be based, with
edges $e_i$ and internal nodes $\rho$, $\rho'$.}
 \label{fig:6taxa}
\end{figure}

To establish identifiability of model parameters on a particular
tree, our argument will require that there be a 6-leaf subtree with
the particular topology shown in Figure \ref{fig:6taxa}. It is easy
to see that any tree with at least 7 leaves contains such a 6-leaf
subtree. (For simplicity, we chose to state Theorem \ref{thm:main}
and its corollary for trees of 7 or more taxa, even though they also
hold for this 6-leaf tree.)

In Section \ref{sec:M9id} the main thread of the proof begins. We
use algebraic arguments built on a theorem of J.~Kruskal
\cite{MR0444690} to determine the covarion Markov matrix
$M=\exp(Rt_9)$ describing the total substitution process over the
central edge $e_9$, of length $t_9$, in the tree of Figure
\ref{fig:6taxa}, up to permutation of the rows and columns. This
part of our argument is not very specific to the covarion model, but
rather applies to more general models provided the Markov matrices
involved satisfy some technical algebraic conditions. We therefore
must show that Markov matrices arising from the covarion model, as
exponentials of a covarion rate matrix, satisfy these technical
conditions, at least for generic parameter choices. Though this fact
is completely plausible, establishing it rigorously requires rather
detailed work, which is completed in Section \ref{sec:covgen}. This
part of our argument is the reason Theorem \ref{thm:main} refers to
identifiability of `generic' parameters and not all parameters, as
well as the reason we require $c\le \kappa$.

Once the Markov matrix on the central edge of the tree is identified
up to row and column permutations, to determine the covarion rate
matrix we must determine the correct row and column orderings, and
take a matrix logarithm. We are able to show there is a unique
ordering of rows and columns that produces a covarion rate matrix in
part by taking advantage of the pattern of zeros that must appear in
such a rate matrix. Other facts about rate matrices, such as the
non-positivity of eigenvalues, also play a role. We obtain an
essential piece of information on the ordering from the known
ordering of bases at the leaves of the tree. All this is the content
of Section \ref{sec:Rid}.

Finally, once we have determined the covarion rate matrix from this
central edge, we use it in Section \ref{sec:edgeid} to determine the
sum of edge lengths between any two leaves in the tree. By standard
arguments, we may then determine the lengths of all individual edges
in the tree, so all parameters have been identified.

\smallskip

Note that the later steps of our arguments are constructive, in that one could apply
them to a specific probability distribution to explicitly recover the parameters producing it.
However, Kruskal's theorem is not constructive; it guarantees a unique set of parameters
but does not indicate a procedure for recovering them. A constructive version
of Kruskal's theorem would give an algorithm for the decomposition of three-dimensional tensors
into minimal sums of rank 1 tensors. This is an interesting but challenging open problem, which would have applications in several other areas of applied mathematics as well.
However, the particular case of Kruskal's theorem we use can also be established by
a longer argument, which we omit, along the lines of the identifiability result in \cite{MR97k:92011}. Using that approach
one obtains an explicit parameter identification procedure that depends on the calculation
of eigenvectors for $c\kappa\times c\kappa$ matrices.

\section{Diagonalizing covarion rate matrices}\label{sec:eig}

We summarize a few basic facts concerning the eigenvectors and eigenvalues of a covarion
rate matrix $R$, under the
hypotheses of the $\Cov(c,\kappa)$ model.

If $R$ is a rate matrix for $\Cov(c,\kappa)$ then it is
time-reversible by assumption. Thus $\diag(\boldsymbol \mu)R$ is
symmetric, and $\diag(\boldsymbol \mu)^{1/2}R\diag(\boldsymbol
\mu)^{-1/2}$ is as well. Therefore $$\diag(\boldsymbol
\mu)^{1/2}R\diag(\boldsymbol \mu)^{-1/2}=C^TBC$$ for some orthogonal
$C$ and real diagonal $B$. Letting $U=C\diag(\boldsymbol
\mu)^{1/2}$, we have
$$R=U^{-1}BU,\ \ \  U^{-1}=\diag(\boldsymbol \mu)^{-1}U^T.$$

If $R$ is class irreducible, then it is irreducible. Thus one of its
eigenvalues is 0 and the others are strictly negative
\cite{hornJohnson85}. We may thus assume
$B=\diag(\beta_1,\beta_2,\dots,\beta_{c\kappa})$, where
$0=\beta_1>\beta_2\ge\dots\ge \beta_{c\kappa}$ for generic $R$.

\smallskip

Note that for the model $\sCov(c,\kappa)$, much more can be said about this diagonalization.
In \cite{GaltJM}, it is shown that the eigenvectors and eigenvalues
for a scaled covarion rate matrix $R$ are related to those of $Q$ and certain modifications of $S$ through a tensor decomposition.

\medskip

We now investigate the implications of the diagonalization of
covarion rate matrices for 2-taxon probability distributions arising
from the model. This will be useful for identifying edge lengths in
Section \ref{sec:edgeid}.

\smallskip

Suppose $R=U^{-1}BU$ is the diagonalization described above. A 2-taxon distribution, arising from edge length $t$, is
described by a $\kappa\times\kappa$ matrix
\begin{align*}
N&=J^T\diag(\boldsymbol\mu)\exp(Rt)J\\
&=J^T\diag(\boldsymbol\mu)U^{-1}\exp(Bt)UJ\\
&=J^TU^T\exp(Bt)UJ\\
&=(UJ)^T\exp(Bt) (UJ). \end{align*}

\smallskip

We formalize this observation with the following lemma.

\begin{lem}\label{lem:2tax}
Let $R$ be a covarion rate matrix for $\Cov(c,\kappa)$. Then $R$
determines a matrix $B=\diag(\beta_1,\dots,\beta_{c\kappa})$ with
$0=\beta_1>\beta_2\ge\dots\ge \beta_{c\kappa}$, and a rank $\kappa$
matrix $K$ of size $c\kappa\times\kappa$ such that the probability
distribution arising from the covarion model with rate matrix $R$ on
a one-edge tree of length $t$ is
$$N= K^T\exp(Bt)K.$$
\end{lem}
\begin{proof}
It only remains to justify that the rank of $K=UJ$ is $\kappa$.
However, since $U$ is non-singular, $\rank K=\rank J=\kappa$.
\end{proof}

\section{ Identifying a Markov matrix on the central
edge}\label{sec:M9id}

The basic identifiability result on which we build our later
arguments is a theorem of J. Kruskal \cite{MR0444690}. (See also
\cite{MR0488592,MR1088949} for more expository presentations.)

For $i=1,2,3$, let $N_i$ be a matrix of size $r\times \kappa_i$,
with $\mathbf n^i_j$ the $j$th row of $N_i$. Let $[N_1,N_2,N_3]$
denote the $\kappa_1\times\kappa_2\times \kappa_3$ tensor defined by
$$[N_1,N_2,N_3]=\sum_{j=1}^r \mathbf n^1_j\otimes \mathbf n^2_j\otimes \mathbf n^3_j.$$
Thus the $(k_1,k_2,k_3)$ entry of $[N_1,N_2,N_3]$ is $\sum_{j=1}^r
\mathbf n^1_j(k_1)\mathbf n^2_j(k_2)\mathbf n^3_j(k_3)$, and this
`matrix triple product' can be viewed as a generalization of the
product of two matrices (with one matrix transposed).

Note that simultaneously permuting the rows of all the $N_i$
(\textit{i.e.}, replacing each $N_i$ by $PN_i$ where $P$ is an
$r\times r$ permutation) leaves $[N_1,N_2,N_3]$ unchanged. Also
rescaling the rows of each $N_i$ so that the scaling factors $c^i_j$
used for the $\mathbf n^i_j$, $i=1,2,3$ satisfy $c^1_jc^2_jc^3_j=1$
(\textit{i.e.}, replacing each $N_i$ by $D_iN_i$, where $D_i$ is
diagonal and $D_1D_2D_3=I$) also leaves $[N_1,N_2,N_3]$ unchanged.
That under certain conditions these are the only changes leaving
$[N_1,N_2,N_3]$ fixed is the essential content of Kruskal's theorem.

To state the theorem formally requires one further definition. For a
matrix $N$, the \emph{Kruskal rank} of $N$ will mean the largest
number $j$ such that every set of $j$ rows of $N$ are independent.
Note that this concept would change if we replaced `row' by
`column,' but we will only use the row version in this paper. With
the Kruskal rank of $N$ denoted by $\rank_K N$, observe that
$$\rank_K N\le \rank N.$$

\smallskip

\begin{thm}(Kruskal)
Let $j_i=\rank_K N_i$. If $$j_1+j_2+j_3\ge 2r+2,$$ then
$[N_1,N_2,N_3]$ uniquely determines the $N_i$, up to simultaneously
permutating and rescaling the rows. That is, if
$[N_1,N_2,N_3]=[N_1',N_2',N_3']$, then there exists a permutation
$P$ and diagonal $D_i$, with $D_1D_2D_3=I$, such that
$N_i'=PD_iN_i$.
\end{thm}
\smallskip

We will apply this result to identify parameters of a stochastic
model with a hidden variable. In phylogenetic terms, the model is
one on a 3-leaf tree, rooted at the central node. A hidden variable
at the central node has $r$ states, and observed variables at the
leaves have $\kappa_1,\kappa_2,\kappa_3$ states respectively. Markov
matrices $M_i$, of size $r\times\kappa_i$, describe transitions from
the state at the central node to those on leaf $i$, with observed
variables conditionally independent given the state of the hidden
variable. For each $i=1,2,3,$ let $\mathbf m^i_j$ denote the $j$th
row of $M_i$. One then checks that the joint distribution for such a
model is given by
$$[\mathbf v; M_1,M_2,M_3]=\sum_{j=1}^r v_j \mathbf m^1_j\otimes \mathbf m^2_j\otimes \mathbf
m^3_j.$$

\smallskip

\begin{cor}\label{cor:kruskal} Suppose $M_i$, $i=1,2,3$, are $r\times\kappa_i$ Markov matrices,
and $\mathbf v=(v_1,\dots,v_r)$ is a row vector of non-zero numbers
summing to 1. Let $j_i=\rank_K M_i$.  If
$$j_1+j_2+j_3\ge 2r+2,$$ then $[\mathbf v; M_1,M_2,M_3]$ uniquely
determines $\mathbf v, M_1,M_2,M_3$ up to permutation. That is,
$[\mathbf v; M_1,M_2,M_3]=[\mathbf v'; M_1',M_2',M_3']$ implies that
there exists a permutation $P$ such that $M_i'=PM_i$ and $\mathbf
v'=\mathbf vP^T$.
\end{cor}
\begin{proof}
This follows from Kruskal's theorem in a straightforward manner,
using that the rows of each Markov matrix $M_i$ sum to 1.
\end{proof}

\smallskip

\begin{rem}
The corollary actually claims identifiability for generic
parameters, where `generic' is used in the sense of algebraic
geometry. To see this, note that for any fixed choice of a positive
integer $j_i$, those matrices $M_i$ whose Kruskal rank is strictly
less than $j_i$ form an algebraic variety. This is because the
matrices for which a specific set of $j_i$ rows are dependent is the
zero set of all $j_i\times j_i$ minors obtained from those rows.
Then, by taking appropriate products of these minors for different
sets of rows we may obtain a set of polynomials whose zero set is
precisely those matrices of Kruskal rank $<j_i$.
\end{rem}

\smallskip

To apply the Corollary of Kruskal's theorem in a phylogenetic
setting, we need one additional definition. Given matrices $N_1$ of
size $r\times s$ and $N_2$ of size $r\times t$, let
$$N=N_1\otimes^{row}N_2$$ denote the $r\times st$ matrix  that is
obtained from row-wise tensor products. That is, the $i$th row of
$N$ is the tensor product of the $i$th row of $N_1$ and the $i$th
row of $N_2$. Although we do not need a specific ordering of the
columns of $N$, we could, for instance, define $N$ by
$N(i,j+s(k-1))=N_1(i,j)N_2(i,k)$.

To interpret this row-wise tensor product in the context of models,
consider a rooted tree with two leaves, and a Markov model with $r$
states at the root, and $\kappa_i$ states at leaf $i$, $i=1,2$. Then
the transition probabilities from states at the root to states at
leaf $i$ are specified by an $r\times\kappa_i$ matrix $M_i$ of
non-negative numbers whose rows add to 1. The matrix
$M=M_1\otimes^{row} M_2$ will also have non-negative entries, with
rows summing to 1. Its entries give transition probabilities from
the $r$ states at the root to the $\kappa_1\kappa_2$ \emph{composite states} at
the leaves, formed by specifying the state at both leaves. Thus this
row tensor operation is essentially what underlies the notion of a
`flattening' of a multidimensional tensor that plays an important
role in \cite{ARgm,ARidtree}.

\medskip

\begin{figure}[h]
\begin{center}
\vskip .1in
\includegraphics[height=.7in]{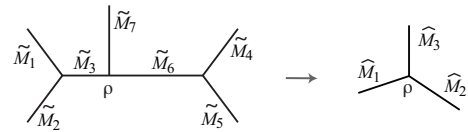}
\end{center}
\caption{Viewing a model on a 5-leaf tree as a model on a 3-leaf
tree.}\label{fig:5taxa}
\end{figure}

Kruskal's result will actually be applied to a model on a 5-leaf
tree, by a method we now indicate. For the 5-leaf tree shown in
Figure \ref{fig:5taxa}, rooted at $\rho$, suppose Markov matrices
$\widetilde M_i$ (not necessarily square) are associated to all
edges to describe transition probabilities of states moving away
from the root.

Then with \begin{align*} \widehat M_1&=\widetilde M_3(\widetilde
M_1\otimes^{row}\widetilde M_2),\\ \widehat M_2&= \widetilde
M_6(\widetilde M_4\otimes^{row} \widetilde M_5),\\ \widehat
M_3&=\widetilde M_7,
\end{align*}  we obtain Markov matrices on a simpler 3-leaf
tree rooted at its central node. Retaining as root distribution the
root distribution $\mathbf v$ at $\rho$, the joint distribution for
this simpler tree is $[\mathbf v; \widehat M_1,\widehat M_2,\widehat
M_3]$. The entries of the distribution for the 5-leaf tree and the
3-leaf tree are of course the same, though one is organized as a
5-dimensional array and the other as a 3-dimensional array. However,
the reorganization into a 3-dimensional array is crucial in allowing
us to apply Kruskal's theorem.

\smallskip

\begin{lem}\label{lem:6krusk}
On the 6-leaf tree of Figure \ref{fig:6taxa} rooted at $\rho$,
consider a Markov model with $r$ states at all internal nodes and
$\kappa$ states at leaves. Let the state distribution at the root be
specified by $\mathbf v$, and Markov matrices $M_i$ describe
transitions on edge $e_i$ directed away from the root, so for
internal edges the $M_i$ are $r\times r$, and on pendant edges are
$r\times\kappa$.

Suppose in addition
\begin{enumerate}
\item[(1)] all entries of both $\mathbf v$ and $\mathbf v'=\mathbf v M_9$ are positive,

\item[(2)] \label{it:mat} the four matrices $M_6(M_4\otimes^{row} M_5)$, $M_9M_6(M_4\otimes^{row}
M_5)$, $M_3(M_1\otimes^{row} M_2)$, and $M_9'M_3(M_1\otimes^{row}
M_2),$
 where $M_9'=\diag(\mathbf v')^{-1}M_9^T\diag(\mathbf v)$, all have rank $r$.

\item[(3)] the Kruskal ranks of $M_7$ and $M_8$ are $\ge 2$.
\end{enumerate}
Then  $M_9$, $M_7$, and $\mathbf v$ are uniquely determined from the
joint distribution, up to permutation. That is, from the joint
distribution we may determine matrices $N_9,N_7$ and a vector
$\mathbf w$ with $N_9=P_1^TM_9P_2$, $N_7=P_1^TM_7$, and $\mathbf
w=\mathbf vP_1$ for some unknown permutations $P_1$ and $P_2$.
\end{lem}

\begin{proof} Note that since the matrices in (2) have rank $r$,
which is equal to the number of their rows, they also have Kruskal
rank $r$.

First consider the 5-leaf subtree where edge $e_8$ has been deleted,
and edges $e_9$ and $e_6$ conjoined. Then by Corollary
\ref{cor:kruskal}, we may determine $\mathbf v P_1$ and the matrices
$P_1^T M_3(M_1\otimes^{row} M_2)$, $P_1^T M_9M_6(M_4\otimes^{row}
M_5)$, and $ P_1^T M_7$ for some unknown permutation $P_1$.

Now reroot the tree of Figure \ref{fig:6taxa} at $\rho'$, using root
distribution $\mathbf v'$ and matrix $M_9'$ on edge $e_9$ (directed
oppositely), without affecting the joint distribution at the leaves.
Having done this, consider the 5-leaf subtree where edge 7 has been
deleted. Another application of the corollary determines $\mathbf
v'P_2$, $P_2^T M_6(M_4\otimes^{row}\overline{} M_5)$, $P_2^T
M_9'M_3(M_1\otimes^{row} M_2)$, and $P_2^TM_8$.

Finally, from the $r\times \kappa^2$ matrices $A=P_1^T
M_9M_6(M_4\otimes^{row} M_5)$ and $B=P_2^T M_6(M_4\otimes^{row}
M_5)$, which by assumption have rank $r$, we may determine the
$r\times r$ matrix $C=P_1^TM_9P_2$: since both $A$ and $B$ have rank
$r$, the equation $A=CB$ uniquely determines $C$.
\end{proof}

\smallskip

Note that for the covarion models, $\mathbf v$ has positive entries by assumption, and since $R$ is time reversible with stationary vector $\mathbf v$, we
will have $\mathbf v'=\mathbf v$ and $M_9'=M_9$. Thus condition (1) will automatically be satisfied in our application of the lemma.

\smallskip

The only potential obstacle to applying Lemma \ref{lem:6krusk} to
the covarion model is that we must know that
assumptions (2) and (3) on the ranks of various products of Markov matrices are
met. While one would certainly suspect that at least for generic
choices of covarion parameters there would be no problem, it is
non-trivial to establish this rigorously. That is the content of the
next lemma.

Let $\{f_1,\dots, f_n\}$ be a finite collection of analytic functions with common domain $D\subseteq\mathbb C^n$.
Recall that the \emph{analytic variety} $V=V(f_1,\dots,f_n)$ is the subset of $D$ on which
all $f_i$ vanish.
In the next lemma we will use the existence of a single point in $D\smallsetminus V$
to conclude that the $V$ is of strictly lower dimension than $D$.
This step may not be
familiar to most researchers in phylogenetics, so we recall a simpler instance.
A powerful theorem concerning analytic functions of a single complex variable
is that if an analytic function $f$
is not identically zero, then any zeros of $f$ in the interior of its domain must be isolated. Equivalently, if there is a single point $z_0$
with $f(z_0)\ne 0$, then the zero set of $f$ is a zero-dimensional subset
of the one-dimensional domain of $f$.
Our argument simply uses a generalization of this fact from the theory of functions of
several complex variables.

\smallskip

\begin{lem} \label{lem:covgen}
Identify the stochastic parameter space $\mathcal S$ of any of the
models $\Cov(c,\kappa)$, $\eCov(c,\kappa)$ or $\sCov(c,\kappa)$ on
the 6-taxon tree of Figure \ref{fig:6taxa} with a full-dimensional
subset of $\mathbb R^L$ so that the parameterization map for the
probability distribution is given by analytic functions.

Let $X\subset \mathcal S$ be the subset on which either at least one
of the four $c\kappa\times \kappa^2$ matrices arising from cherries,
\begin{gather*}
\exp(Rt_3)(\exp(Rt_1)J\otimes^{row}\exp(Rt_2)J),\\
\exp(R(t_3+t_9))(\exp(Rt_1)J\otimes^{row}\exp(Rt_2)J),\\
\exp(Rt_6)(\exp(Rt_4)J\otimes^{row}\exp(Rt_5)J),\\
\exp(R(t_6+t_9))(\exp(Rt_4)J\otimes^{row}\exp(Rt_5)J),
\end{gather*}
has rank $<c\kappa$, or at least one of the two matrices
$$\exp(Rt_7)J,\ \ \ \exp(Rt_8)J$$ on the pendant edges $e_7$, $e_8$ has Kruskal rank $< 2$.
Then if $c\le \kappa$, the set $X$
 is a proper analytic
subvariety of $\mathcal S$, and hence of dimension $<L$.
\end{lem}

\begin{proof}
For our argument, it will be convenient to extend the set of
allowable edge lengths from $t_i>0$ to a larger set including $t_i=0$. Once the claim is
established allowing zero-length edges, we may restrict to
positive-length edges (as is needed in other parts of our paper).
This is simply because the original and extended parameter spaces described here
have the same dimension, so the intersection of a proper analytic subvariety of the extended
parameter space with the smaller parameter space must also be a proper subvariety.

Consider first the edges $e_1,e_2,e_3,e_7$ in the tree of Figure
\ref{fig:6taxa}. In Section \ref{sec:covgen} below it will be shown
that when $c\le \kappa$ there is at least one choice of  a rate matrix $R$ for
$\sCov(c,\kappa)$, and edge lengths $t_1>0$, $t_2=0$, $t_3=0$, $t_7>0$ so that $\exp(Rt_3)(\exp(Rt_1)J\otimes^{row}\exp(Rt_2)J)$
has rank $c\kappa$ and $\exp(Rt_7)J$ has Kruskal rank $\ge2$.
Assuming this result for now, by in addition choosing
$$t_9=0,\ t_8=t_7,\ t_6=t_3,\ t_5=t_2,\ t_4=t_1$$
we have found at least one parameter choice for $\sCov(c,\kappa)$
that does not lie in $X_{\sCov}$.

Since the same $R$ and $\{t_i\}$ arise from parameters for
$\eCov(c,\kappa)$, respectively $\Cov(c,\kappa)$, we have also found
at least one parameter choice for these models that does not lie in
$X_{\eCov}$, respectively $X_{\Cov}$.

Now observe that the set of parameters for which any one of the four
specified $c\kappa\times\kappa^2$ matrices has rank $<c\kappa$ is
the zero set of a collection of analytic functions. Such functions
can be explicitly constructed by composing the parameterization map
for each matrix with the polynomial functions expressing the
$c\kappa\times c\kappa$ minors. Similarly, the set of parameters for
which a pendant edge matrix fails to have Kruskal rank $\ge 2$ is
the simultaneous zero set of a collection of analytic functions
built from the composition of the parameterization of that matrix
with the $2\times 2$ minors. Thus the set $X$ is the union of
analytic varieties, and hence itself an analytic variety. This set cannot
be the entire parameter space, since we have found one point that
lies outside it. Therefore $X$ is a proper analytic subvariety, as
claimed. As such, it is of dimension strictly less than $L$.
\end{proof}

\smallskip

For all covarion parameters outside the set $X$ of Lemma
\ref{lem:covgen}, we may apply Lemma \ref{lem:6krusk} and identify
$M=P_1^T\exp(Rt_9)P_2$ and $\boldsymbol \nu =\boldsymbol \mu P_1$
for some unknown permutations $P_1,P_2$. As $X$ is of lower
dimension than the parameter space, it has Lebesgue measure 0. Thus
for generic covarion parameters we may identify $M$ and $\boldsymbol
\nu$.

\section{Construction of scaled covarion parameters with certain properties}\label{sec:covgen}

In this section the particular parameter choice needed in the proof
of Lemma \ref{lem:covgen} is constructed. We thus consider only the model
$\sCov$, with the parameters $Q$, $S$, and $\{r_i\}$ as described in Section \ref{sec:param}, and
$R$ as given by equation (\ref{eq:RsCov}). We seek values of these parameters and of $t_1,t_7>0$ so that
$\exp(Rt_1)J\otimes^{row} J$ has rank $c\kappa$ and
$\exp(Rt_7)J$ has Kruskal rank at least 2.
Note that since $\exp(Rt_1)J\otimes^{row} J$ is $c\kappa\times \kappa^2$, it may only have the desired
rank when $c\le \kappa$.
\smallskip

One might first
consider taking $t_1=0$, so
$$\exp(Rt_1)J\otimes^{row} J=J\otimes^{row}J.$$ However this $c\kappa\times \kappa^2$ matrix
has rank $\kappa<c\kappa$. Similarly, taking $t_7=0$, so $\exp(Rt_7)J=J$, fails to produce a matrix of Kruskal rank at least 2. Thus we must do more work to find the needed example.
Our first step is to establish the following.

\smallskip

\begin{lem}
Suppose that for each $j\in [\kappa]$, the vectors appearing as the $j$th rows of the matrix powers $Q^m$, $m=1,\dots,
{c-1}$ are independent. Then there exist $t_1,t_7>0$ such that $\exp(Rt_1)J\otimes^{row}J$
has rank $c\kappa$ and $\exp(Rt_7)J$ has Kruskal rank at least 2.
\end{lem}

\begin{proof}
We first show the existence of such a $t_1$. Let $M=M(t)=\exp(Rt)J$. Because of the
specific form of $J$, it is easy to see that any dependency
relationship between the rows of $M \otimes^{row}J$ is equivalent
to $\kappa$ separate dependency relationships between rows of $M$.
Specifically, the rows of $M\otimes^{row}J$ are independent if,
and only if, for each $j\in[\kappa]$ the set of the $c$ rows of $M$ with index
$(i,j)$, $i\in [c]$, are independent.

Letting $X_j(t)$ denote the $c\times \kappa$ submatrix of $M(t)$ consisting of the $(i,j)$ rows,
we claim that some $c\times c$ minor of $X_j(t)$
is non-zero for all but a discrete set of values of $t$. Since there are only finitely many $j$
to consider, this implies the existence of the desired $t_1$.

Fixing $j$, for notational ease let $$X_j(t)=\begin{pmatrix} \mathbf x_1(t)\\ \vdots\\\mathbf x_c(t)\end{pmatrix},\ \ x(t)=\det \begin{pmatrix} \bar{\mathbf x}_1(t)\\ \vdots\\ \bar{\mathbf x}_c(t)\end{pmatrix}$$ where
the bar denotes projection onto some choice of $c$ coordinates, to be specified later, so
that $x(t)$ is a specific $c\times c$ minor of $X_j(t)$.

Since $x(t)$ is an analytic function, to establish that it is non-zero except at a discrete set of points, it is enough to show it is not identically zero. Now $x(t)$ is easily evaluated only at $t=0$, and unfortunately $x(0)=0$ since $\mathbf x_i(0)$ is the standard basis vector $\mathbf e_j$ for all $i$. We will, however, show $x(t)$ is not identically zero by showing the derivative  $x^{(n)}(0)$ is non-zero for $n=c(c-1)/2$.

To obtain information on the derivatives $\mathbf x_i^{(l)}(0)$, observe that $M(t)$ is the solution to the initial value problem $M'=RM$, $M(0)=J$. Thus $\mathbf x_i^{(l)}(0)$ is the $(i,j)$ row of $R^{l}J$.
Moreover, since $S\mathbf 1_c^T=\mathbf 0$,
\begin{align*}R^lJ &=( \diag(r_1,r_2,\dots, r_c)\otimes Q+S\otimes I_\kappa)^l (\mathbf 1_c^T\otimes I_\kappa)\\
&= \diag(r_1,r_2,\dots, r_c)^l \mathbf 1_c^T \otimes Q^l +\sum_{m=1}^{l-1} \mathbf y_{l,m}^T\otimes Q^m,\end{align*}
for some vectors $\mathbf y_{l,m}$. Thus, for $l\ge 1$, $\mathbf x_i^{(l)}(0)$ is a linear combination of the $j$th rows of $Q^m$, $1\le m\le l$,
where the $j$th row of $Q^l$ appears with coefficient $r_i^l$.

Now with $n=c(c-1)/2$,
\begin{equation}x^{(n)}(0)= \sum _{\lambda=(n_1,\dots, n_c) } m_\lambda \det \left( \bar{ \mathbf x}_1^{(n_1)}(0),\, \dots \, ,\bar{ \mathbf x}_c^{(n_c)}(0) \right),\label{eq:detex}
\end{equation}
 where
the summation is over non-negative integer solutions to $n_1+\cdots+n_c=n$ and
$m_\lambda= \binom{n}{n_1,\dots,n_c}$ is a multinomial coefficient.
Letting $\mathbf z_0=\mathbf e_j$ and $\mathbf z_i$ be the $j$th row of $Q^i$ for $i\ge 1$, we have shown that $\mathbf x_i^{(l)}(0)$ lies in the span of $\{\mathbf z_i\}_{i=0}^{l}$ for all $l\ge 0$. This implies that any summand in
equation (\ref{eq:detex}) must vanish if more than $l+1$ of the $n_i$ satisfy $n_i\le l$,
since in that case the rows in the determinant are dependent. But
$n=c(c-1)/2=0+1+\cdots+(c-1)$, hence non-zero terms can arise only when $\lambda$ is a permutation
of $(0,1,\dots,c-1)$.

With $S_c$ denoting the permutations of $(1,\dots,c)$, and
$m=m_{(0,1,\dots,c-1)}$,
\begin{align*}x^{(n)}(0)& = m \sum _{\mu \in S_c}  \det\left ( \bar{ \mathbf x}_1^{(\mu^{-1}(1)-1)}(0),\, \dots \, ,\bar{ \mathbf x}_c^{(\mu^{-1}(c)-1)}(0)\right )\\
&=m \sum _{\mu \in S_c} \operatorname{sgn}(\mu) \det\left ( \bar{ \mathbf x}_{\mu(1)}^{(0)}(0),\, \dots \, ,\bar{ \mathbf x}_{\mu(c)}^{(c-1)}(0) \right).
\end{align*}
But with
$Z=(\mathbf z_0^T,\dots,\mathbf z_{c-1}^T)^T$, we have shown
$$\begin{pmatrix} { \mathbf x}_{\mu(1)}^{(0)}(0)\\\vdots \\ { \mathbf x}_{\mu(c)}^{(c-1)}(0)
\end{pmatrix} =L_\mu Z$$
where $L_\mu$ is a $c\times c$ lower triangular matrix with diagonal entries $L_{i,i}=r_{\mu(i)}^{i-1}$.
By hypothesis, all rows of $Z$ except the first form an independent set, and since $Q^l \mathbf 1_c^T=\mathbf 0$ for $l\ge 1$ while $\mathbf z_0 \mathbf 1_c^T=1$, the first row is not in the span of the others. Thus $Z$ has rank $c$, and some choice of $c$ of its columns are independent. Specifying that the bar over a matrix or row vector designates a projection onto these column coordinates yields
$$\begin{pmatrix} \bar{ \mathbf x}_{\mu(1)}^{(0)}(0)\\\vdots \\ \bar{ \mathbf x}_{\mu(c)}^{(c-1)}(0)
\end{pmatrix} =L_\mu \bar{ Z},$$ so
$$\det\left ( \bar{ \mathbf x}_{\mu(1)}^{(0)}(0),\dots, \bar{ \mathbf x}_{\mu(c)}^{(c-1)}(0)\right )
=\left (\prod_{i=1}^c r_{\mu(i)}^{i-1} \right )\det(\bar{ Z}).$$
Since $\det(\bar Z)\ne 0$, to see that $x^{(n)}(0)\ne 0$ it is enough to show
$$\sum _{\mu \in S_c} \operatorname{sgn}(\mu) \prod_{i=1}^c r_{\mu(i)}^{i-1} \ne 0$$
But the left hand side is a Vandermonde determinant, and since the $r_i$ are distinct, it does not vanish. Thus the desired $t_1$ exists.

For the existence of $t_7$, consider the $(i_1,j_1)$ and $(i_2,j_2)$ rows of $\exp(Rt)J$. If $j_1\ne j_2$,
then these rows are independent when $t=0$, hence for all $t$ except a discrete set.
If $j_1=j_2$, then the two rows are rows of $X_{j_1}(t)$, and thus independent for all but a discrete set
of $t$ by our work above. Since there are only finitely many pairs to consider, for all but a discrete
set of values we find $\exp(Rt)J$ has Kruskal rank $\ge 2$.
\end{proof}

\smallskip

The existence of rate matrices $Q$ satisfying the hypotheses of the last lemma
is a consequence of the following one.

\smallskip

\begin{lem}\label{lem:Qpower}
Suppose a $\kappa\times\kappa$ rate matrix $Q$ has at least $c$ distinct eigenvalues and its
right eigenvectors can be chosen to have all non-zero entries.
Then
for each $j\in[\kappa]$ the vectors appearing as the $j$th rows of $Q^l$, $l=0,\dots,c-1$, are independent.
\end{lem}
\begin{proof}
Let $Q=UDU^{-1}$ be a diagonalization of $Q$. Then with $\mathbf u_j$ denoting the $j$th row of $U$,
the $j$th row of $Q^l$ is $\mathbf u_jD^l U^{-1}$. To show these rows are independent, it is
enough to show the $\mathbf u_jD^l$, $l=0,\dots,c-1$ are independent, or even that the projections
of these vectors onto some choice of $c$ coordinates are independent. By choosing to project onto
$c$ coordinates corresponding to distinct diagonal entries of $D$, we may reduce to the case
where $D$ is $c\times c$ with distinct diagonal entries and the vectors $\mathbf u_j\in \mathbb C^{c}$ have all non-zero entries.

But if $W$ is the $c\times c$ matrix whose $l$th row is $\mathbf u_jD^{l-1}$, then $W=V\diag(\mathbf u_j)$ where $V$ is a Vandermonde matrix
constructed from the diagonal entries of $D$. By our assumptions, both $V$ and $\mathbf \diag(\mathbf u_j)$ have non-zero determinants, so $W$ does as well. Thus the rows of $W$ are independent.
\end{proof}

\smallskip

To see a $Q$ satisfying the hypotheses of Lemma \ref{lem:Qpower} exists, let $$Q_0=\frac 1{\kappa(\kappa-1)}\left ( \mathbf 1_\kappa^T\mathbf 1_\kappa-\kappa I_\kappa\right )$$ be a generalized
Jukes-Cantor matrix of size $\kappa$, all of whose off-diagonal entries are equal, which has stationary vector $\mathbf 1_\kappa$. The eigenspaces
of $Q_0$ are the span of $\mathbf 1_\kappa$ and its orthogonal complement. For a diagonalization $Q_0=UD_0U^{-1}$ we can thus chose $U$ to be an orthogonal matrix all of whose entries are non-zero.
(For instance, when $\kappa=4$ we may choose $U$ to be a Hadamard matrix.)
Since
$D_0$ has repeated diagonal entries, perturb the non-zero entries slightly to obtain a diagonal matrix $D$ without repetitions, and let $Q=UDU^{-1}$. Since $Q$ also has $\mathbf 1_\kappa$ as its
stationary distribution, and since $Q$ is symmetric, it is a rate matrix of the sort needed.

\smallskip

Choosing such a $Q$ and any $S$ and distinct $r_i$ for the $\sCov$ parameters gives a particular choice of scaled covarion parameters $Q$, $S$, $\{r_i\}$ such that there exists a $t_1>0$
where $\exp(Rt_1)J\otimes J$ has
rank $c\kappa$, and a $t_7\ge 0$ such that $\exp(Rt_7)J$ has Kruskal rank at least 2.

Thus Lemma \ref{lem:covgen} is fully established.

\section{Identifying the covarion rate matrix $R$}\label{sec:Rid}

The next goal is to use $\boldsymbol\nu=\boldsymbol \mu P_1$ and
$M=P_1^T\exp(Rt_9)P_2$, as identified in Section \ref{sec:M9id}
through Lemmas \ref{lem:6krusk} and \ref{lem:covgen}, to determine
the covarion root distribution $\boldsymbol\mu$ and the covarion
rate matrix $R$. It is of course enough to determine $Rt_9$, where
$t_9> 0$ is the edge length, and then use the required normalization
of $R$.

Let us assume $\boldsymbol\nu$ has its entries in non-increasing
order. (This can be achieved by multiplying $\boldsymbol \nu$ on the
right by some permutation $P$, and $M$ on the left by $P^T$, thereby
changing the unknown $P_1$.) Now since
$\diag(\boldsymbol\mu)\exp(Rt)$ is symmetric, and
$\diag(\boldsymbol\nu)= P_1^T\diag(\boldsymbol \mu)P_1$, one can
verify that $\diag(\boldsymbol\nu) M P_2^TP_1$ is symmetric as well.
This shows there is at least one reordering of the columns of $M$
that results in $\diag(\boldsymbol\nu) M$ being symmetric. Assume
some such ordering of the columns of $M$ has been chosen to ensure
this symmetry.

If $\boldsymbol\nu$ (equivalently, $\boldsymbol\mu$) has no repeated
entries, these choices have uniquely determined an ordering to the
rows and columns of $M$, and forced $P_2=P_1$. To see this, note the
rows of $M$ have a fixed correspondence to entries of
$\boldsymbol\nu$, which have a unique decreasing ordering. For the
columns, note that the symmetry of $\diag(\boldsymbol \nu)M$ and the
fact that $\mathbf 1_{c\kappa}M^T=\mathbf 1_{c\kappa}$ implies
$\boldsymbol\nu M=\boldsymbol\nu$. However, if the columns of $M$
are permuted by $P$, then $\boldsymbol\nu MP =\boldsymbol\nu P\ne
\boldsymbol\nu$. We therefore can conclude
$\boldsymbol\nu=\boldsymbol\mu P_1$ and $M=P_1^T \exp(Rt_9)P_1$ for
some unknown permutation $P_1$.

Since $\boldsymbol\nu$ may have repeated entries, the above argument
only holds for generic choices of parameters. In order to avoid
introducing any generic conditions other than those already arising
from the application of Kruskal's theorem, we give an alternate
argument using the following lemma.

\smallskip

\begin{lem}\label{lem:symineq}
Suppose that a matrix $M$ has a factorization of the form $M=PW^TZW$ for some real symmetric positive-definite $m\times m$
matrix $Z$, real $m\times n$ matrix $W$ of rank $n$, and $n\times n$
permutation $P$. Then $P$ is uniquely
determined by $M$.
\end{lem}
\begin{proof}
The matrix $Z$ defines an inner product on $\mathbb R^m$, and if
$\boldsymbol w_i$ denotes the $i$th column of $W$, then the $i,j$
entry of the symmetric matrix $N=W^TZW$ is
$$ \langle \boldsymbol w_i, \boldsymbol
w_j\rangle_Z=\boldsymbol w_i^T Z \boldsymbol w_j.$$ But for any
inner product, if $\boldsymbol x\ne \boldsymbol y$ then
$$\langle \boldsymbol x,\boldsymbol x\rangle +\langle \boldsymbol y,\boldsymbol y\rangle >
2\langle \boldsymbol x, \boldsymbol y\rangle.$$  Now the matrix $W$
has distinct columns since it has rank $n$. Thus the entries of $N$
satisfy
\begin{equation} n_{ii} +n_{jj}>2
n_{ij}.\label{eq:bineq}
\end{equation}

Suppose for some permutations $P_1,P_2$ the matrices $N_1=P_1^TM$
and $N_2=P_2^TM$ are both symmetric, and have entries satisfying the
inequalities (\ref{eq:bineq}). Note also that $N_1$ and $N_2$ have
the same set of rows.

Consider first the largest entry (or entries, in case of ties) of
$N_1$ and $N_2$. Because the inequality in (\ref{eq:bineq}) is
strict, a largest entry cannot appear off the diagonal. Thus the row
(or rows) of $N_1$ and $N_2$ containing the largest entry (or
entries) must occur in the same positions. Since the same argument
applies to the submatrices obtained from the $N_i$ by deleting the
rows and columns with the largest entries, repeated application
shows $N_1=N_2$. Thus $P_1=P_2$.
\end{proof}

\smallskip

\begin{cor}
Suppose $\boldsymbol\nu$, $M$ are of the form
$$\boldsymbol\nu =\boldsymbol\mu P_1,\ \ \ \   M=P_1^T\exp(Rt)P_2,$$
for some covarion rate matrix $R$ with stationary vector
$\boldsymbol \mu$, permutations $P_1,P_2$, and scalar $t$. Then
$P_1^TP_2$ is uniquely determined.
\end{cor}
\begin{proof}
Apply Lemma \ref{lem:symineq} to $\diag(\boldsymbol \nu)M$, with $P=P_1^TP_2$, $W=P_2$, and $Z=\diag(\boldsymbol \mu)\exp(Rt)$.
\end{proof}

\smallskip

As a consequence of this corollary, after multiplying $M$ on the
right by $(P_1^TP_2)^T$ we may now assume we have
$$\boldsymbol\nu=\boldsymbol\mu P,\ \ \ \ M=P^T \exp(Rt) P$$ for some
(unknown) permutation $P$. But then $M=\exp(P^TRPt)$, and since this
matrix is diagonalizable with positive eigenvalues, $P^TRPt$ is
determined by applying the logarithm to its diagonalization.

Now $P^TRPt$ is simply a rescaled version of $R$ with the same
permutation applied to rows and columns. Thus there exists at least
one simultaneous permutation of the rows and columns of $P^TRPt$ which yields a
rescaled covarion rate matrix. However, we do not yet know if there is
a unique such permutation, or a unique such covarion rate matrix.

One might suspect that the pattern of zero entries in the
off-diagonal blocks of a covarion rate matrix should allow the
(almost) unique determination of $Rt$ from this permuted form. This
is the content of the following lemma.

\smallskip

\begin{lem}\label{lem:GTRconj}
Let $R_1,R_2$ be rate matrices for $\Cov(c,\kappa)$, with $R_1$
class irreducible, as defined in Section \ref{sec:param}. Suppose
for permutations $P_1,P_2$, and scalars $t_1,t_2>0$, that
$$P_1^TR_1P_1t_1=P_2^TR_2P_2t_2.$$
If $c\ne \kappa$ then $t_1=t_2$, and $P=P_1P_2^T$ can be expressed
as $P=\widehat P\otimes\widetilde P$ for some $c\times c$
permutation $\widehat P$ and $\kappa\times \kappa$ permutation
$\widetilde P$. Thus $R_1$ can be determined up to application of a
permutation of the form $\widehat P\otimes\widetilde P$.

If $R_1, R_2$ are rate matrices for either $\sCov(c,\kappa)$ or
$\eCov(c,\kappa)$, then the same result holds for all $c$.
\end{lem}

\medskip

Note that a permutation of the form $\widehat P\otimes\widetilde P$
can be viewed as a permutation of classes by $\widehat P$, and a
simultaneous permutation of bases within all classes by $\widetilde
P$.

\begin{proof} Using the normalization of $R_1$ and $R_2$, it is trivial to see that $t_1=t_2$.
Conjugating by $P_2$, we obtain $P^TR_1P=R_2$.

Let $N$ be a matrix of the same size as $R_1$, with entry 1
(respectively, 0) wherever the corresponding entry of $R_1$ is
positive (respectively non-positive). Let $G_1=G(R_1)$ be the
(undirected) graph whose adjacency matrix is $N=N^T$. Thus the
vertices of $G_1$ are labeled by the elements of
$[c]\times[\kappa]$, the indices corresponding to rows and columns
of $R_1$, and an edge joins vertices $i$ and $j$ exactly when
$R_1(i,j)> 0$ (or, equivalently, when $R_1(j,i)> 0$). $G_1$ is the
`communication graph' of $R_1$, expressing which instantaneous state
changes can occur.

By assumptions on $R_1$, for each class $i$ with $Q_i\ne 0$, the
vertices labeled $(i,j)$, $j\in[\kappa]$, corresponding to all
states in class $i$, form a clique (\textit{i.e.}, the subgraph on
these vertices is a complete graph) of size $\kappa$. Moreover,
these cliques are each maximal, since any vertex $(i',j')$ outside
of the clique has $i'\ne i$ and is connected to at most one vertex
in the clique, namely $(i,j')$, which has the same base but
different class.

\smallskip

Suppose first that $c\ne \kappa$. In this case we show there are no
other maximal cliques of size $\kappa$. To this end, suppose a
vertex labeled $(i,j)$ is in some other maximal clique $\mathcal C$
of size $\kappa$. The only vertices adjacent to it outside of its
class correspond to the same base $j$. Thus $\mathcal C$ must
contain at least one of these, say $(k,j)$ where $k\ne i$. As the
$(k,j)$ vertex and any $(i,l)$ vertex cannot be in a common clique
if $j\ne l$, $\mathcal C$ must contain only vertices corresponding
to base $j$. As there are $c\ne\kappa$ of these, they cannot form a
clique of size $\kappa$.

Now if we similarly construct $G_2=G(R_2)$, the statement
$P^TR_1P=R_2$ means there is a graph isomorphism from $G_1$ to
$G_2$, obtained by relabeling vertices according to the permutation
$P$. As such an isomorphism must take maximal cliques to maximal
cliques, we see that $P$ must map all states in an $R_1$ class with
$Q_i\ne 0$ to all states in an $R_2$ class with $Q_j\ne0$. (As the
covarion model allows at most one class with $Q_i=0$, this also
means that if either $R_i$ has a class with $Q_i=0$, then so does
the other, and these classes must also be mapped to one another.)

This implies $P$ has the following structure: Partition $P$ into a
$c\times c$ matrix of $\kappa\times\kappa$ blocks, corresponding to
classes. All blocks of $P$ are zero, except for one block in each
row and column. Let $\widehat P$ be the $c\times c$ permutation
matrix with 1s in positions corresponding to those non-zero blocks.
The non-zero blocks of $P$ are also $\kappa\times\kappa$ permutation
matrices.

We next claim that the non-zero $\kappa\times \kappa$ blocks in $P$
are all identical. To see this, consider how $P$ acts on a non-zero
off-diagonal block $S_{i_1i_2}$ of $R_1$ through the formula
$P^TR_1P$: the resulting block has the form $\widetilde P_1^T
S_{i_1i_2} \widetilde P_2$ where $\widetilde P_1$ and $\widetilde
P_2$ are two of the $\kappa\times\kappa$ permutations appearing as
blocks of $P$. But this must equal the corresponding block of $R_2$,
which is diagonal. Thus if all diagonal entries of $S_{i_1i_2}$ are
non-zero then $\widetilde P_1^T \widetilde P_2=I_\kappa$, so
$\widetilde P_1=\widetilde P_2$. The class irreducibility of $R_1$
ensures that we obtain enough such equalities to see that all
$\widetilde P_i$ are equal to some common $\kappa\times\kappa$
permutation $\widetilde P$. Thus $P=\widehat P\otimes \widetilde P$.

\smallskip

Now for the models $\sCov$ and $\eCov$ consider the case of
$c=\kappa$. In this case, maximal cliques corresponding to either a
fixed base or a fixed class have the same cardinality, but there can
be no other maximal cliques. Unless the graph isomorphism from $G_1$
to $G_2$ maps some fixed-base clique to a fixed-class clique, our
earlier argument applies.

We therefore suppose that the  base $j$ clique is mapped to the
class $i$ clique, and argue toward a contradiction. This means $P$
maps vertices in $G_1$ labeled $(k,j)$ for $k=1,\dots,c$ to
vertices labeled $(i,l)$ for $l=1,\dots,\kappa$ in $G_2$. As a result,
every other fixed-base clique in $G_1$ must also map to a
fixed-class clique in $G_2$, since all the fixed-base cliques of
$G_2$ include some $(i,l)$.

But the formula $P^TR_1P=R_2$ implies that each diagonal block of
$R_2$ must have as its $\kappa^2-\kappa$ off-diagonal entries the
$\kappa^2-\kappa$ values $s_{i_1i_2}\ne 0$ which appear in the
off-diagonal blocks of $R_1$. But this is impossible, since the
base-change matrices $Q_i$ of $R_2$ are assumed not to be equal.
\end{proof}

\smallskip

We now have determined $R$ and $\boldsymbol \mu$ up to separate
permutations $\widetilde P$ of the bases and $\widehat P$ of the
classes. The ambiguity expressed by $\widehat P$ cannot be removed,
as permuting classes has no effect on the distributions defined by
the model. Our next step is to use information on the ordering of
the bases obtained at the leaves of the tree in order to determine
$\widetilde P$.

\smallskip

Let $P^TM_7$ denote the $c\kappa\times \kappa$ matrix, which was
determined via Lemma \ref{lem:6krusk}, describing permuted
transition probabilities on edge $e_7$ of the tree of Figure
\ref{fig:6taxa}. Assuming $P=\widehat P\otimes\widetilde P$ by
previous steps in our analysis, $(\widehat P \otimes \widetilde
P)^T\exp(R t_7)J$ is known.

\smallskip

\begin{lem}\label{lem:PPid}
Suppose $W=P^T\exp(R t)J$ for some
permutation $P=\widehat P \otimes \widetilde P$, covarion rate matrix $R$,
and scalar $t$. Then $\widetilde P$ is uniquely determined.
\end{lem}

\begin{proof}
Consider the $\kappa\times\kappa$ matrix, determined by known
information,
\begin{align*} J^T \diag(\boldsymbol \nu)
W&=J^TP^T\diag(\boldsymbol \mu)PP^T\exp(R t_7)J\\
&=(\mathbf 1_c\otimes I_\kappa)(\widehat P^T\otimes\widetilde P^T)
\diag(\boldsymbol\mu)\exp(R t_7)J\\
&=(\mathbf 1_c\widehat P^T\otimes I_\kappa\widetilde P^T)
\diag(\boldsymbol\mu)\exp(R t_7)J\\
&=(\mathbf 1_c\otimes \widetilde P^T)
\diag(\boldsymbol\mu)\exp(R t_7)J\\
&=\widetilde P^T (\mathbf 1_c\otimes I_\kappa)
\diag(\boldsymbol\mu)\exp(R t_7)J\\
&=\widetilde P^T N,
\end{align*}
where  $N=J^T\diag(\boldsymbol\mu)\exp(R t_7)J$. From Lemma
\ref{lem:2tax}, we also have that
$$
N=  K^T \exp(B t_7) K$$ where $B$ is real diagonal and $K$ has rank
$\kappa$. We may thus apply Lemma \ref{lem:symineq} to the product $$J^T\diag(\nu)W=\widetilde P^T K^T \exp(Bt_7)K$$ to determine
$\widetilde P$.
\end{proof}

Thus for generic parameters, $R$ and $\boldsymbol \mu$ are
determined uniquely, up to the permutation $\widehat P$ of classes.

\smallskip
\begin{rem}
That the restriction $c<\kappa$ is necessary for the $\Cov$ model in
Lemma \ref{lem:GTRconj} can be easily seen. For example, with $\kappa=c=2$, the two rate matrices \begin{align*}R&=\frac 1{14}\begin{pmatrix} -5 &3& 2& 0\\
3&-4 &0 &1\\2&0&-3&1\\0& 1 &1& -2\end{pmatrix},\\
 R'&=\frac 1{14}\begin{pmatrix} -5 &2& 3& 0\\
2&-3 &0 &1\\3&0&-4&1\\0& 1 &1& -2\end{pmatrix}
\end{align*}
are related by exchanging rates and classes. Note further that both $R$ and $R'$ have $\frac 14\mathbf 1_4$ as their
stationary distribution, so they lead to the same observed distribution at a single leaf. Moreover,
they lead to the same set of observable distributions at two leaves when one considers all possible edge lengths $t\ge 0$.
Thus one cannot use the observed
distribution at one or two leaves to distinguish between distributions
arising from these two rate matrices.

Of course one might next attempt to use observed joint distributions at multiple leaves to
distinguish these parameters, or introduce additional generic conditions to
obtain identifiability of numerical $\Cov$ parameters even when $c=\kappa$. As we have
not pursued these directions, we do not claim identifiability
fails for generic parameters in this case, but only that the arguments given above
do not establish it.
\end{rem}

\section{Identifying edge lengths}\label{sec:edgeid}

As $R$ is now known, all that remains is to determine edge lengths.
By simple and well-known arguments \cite{MR2060009}, these can be
determined from knowing total distances between leaves of the tree.
Thus the determination of all edge lengths is established by the
following.

\smallskip

\begin{lem}
Fix a covarion rate matrix $R$, of size $c\kappa\times c\kappa$.
Suppose a $\kappa\times \kappa$ matrix $N$ is in the image of the
resulting covarion model on a 2-taxon tree, with edge length $t$.
Then $N$ uniquely determines $t$.
\end{lem}

\begin{proof} From Lemma \ref{lem:2tax}, we have that
\begin{equation*} N=K^T\exp(Bt) K,
\end{equation*}
where $B=\diag(\beta_1,\dots,\beta_{c\kappa})$,
$0=\beta_1>\beta_2\ge\dots \ge \beta_{c\kappa}$ and $K$ is a real
$c\kappa\times \kappa$ matrix, of rank $\kappa$. Furthermore, since
$R$ is known, so are all $\beta_i$ and $K$.

With $K=(k_{ji})$ and $N=(n_{ij})$, this implies the diagonal
entries of $N$ are
\begin{equation}\label{eq:Niisum}
 n_{ii} =\sum_{j=1}^{c\kappa}
k_{ji}^2\exp(\beta_jt).
\end{equation}
As the $k_{ji}$ are real numbers and all $\beta_i$ are non-positive,
each term in this formula is a non-increasing function of $t$. Thus
$n_{ii}=n_{ii}(t)$ is a non-increasing function of $t$. If we show
that for some $i$  the function $n_{ii}(t)$ is strictly decreasing,
then from any value of $n_{ii}$ we may determine $t$. But to
establish that some $n_{ii}$ is strictly decreasing, we need only
show there exists some $i$ and some $j>1$ such that $k_{ji}\ne 0$,
so that at least one term in equation (\ref{eq:Niisum}) is a
strictly decreasing function. However, as $K$ has rank $\kappa>1$,
we cannot have $k_{ji}=0$ for all $j>1$.
\end{proof}

\bibliographystyle{plain}

\bibliography{Phylo}

\def\cprime{$'$}
\begin{thebibliography}{10}

\bibitem{AllmanAneRhodes08}
Elizabeth~S. Allman, C\'ecile An\'e, and John~A. Rhodes.
\newblock Identifiability of a {M}arkovian model of molecular evolution with
  gamma-distributed rates.
\newblock {\em Advances in Applied Probability}, 40(1):229--249, 2008.
\newblock {\tt arXiv:0709.0531}.

\bibitem{ARidtree}
Elizabeth~S. Allman and John~A. Rhodes.
\newblock {The identifiability of tree topology for phylogenetic models,
  including covarion and mixture models}.
\newblock {\em J. Comput. Biol.}, 13(5):1101--1113, 2006.
\newblock {\tt arXiv:q-bio.PE/0511009}.

\bibitem{ARGMI}
Elizabeth~S. Allman and John~A. Rhodes.
\newblock Identifying evolutionary trees and substitution parameters for the
  general {M}arkov model with invariable sites.
\newblock {\em Math. Biosci.}, 211(1):18--33, 2008.
\newblock {\tt arXiv:q-bio.PE/0702050}.

\bibitem{ARgm}
Elizabeth~S. Allman and John~A. Rhodes.
\newblock Phylogenetic ideals and varieties for the general {M}arkov model.
\newblock {\em Adv. in Appl. Math.}, 40(2), 2008.
\newblock {\tt arXiv:math.AG/0410604}.

\bibitem{MR97k:92011}
Joseph~T. Chang.
\newblock Full reconstruction of {M}arkov models on evolutionary trees:
  identifiability and consistency.
\newblock {\em Math. Biosci.}, 137(1):51--73, 1996.

\bibitem{FM}
Walter~M. Fitch and Etan Markowitz.
\newblock An improved method for determining codon variability in a gene and
  its application to the rate of fixation of mutations in evolution.
\newblock {\em Biochemical Genetics}, 4:579--593, 1970.

\bibitem{Galt}
Nicolas Galtier.
\newblock Maximum-likelihood phylogenetic analysis under a covarion-like model.
\newblock {\em Mol. Biol. Evol.}, 18(5):866--873, 2001.

\bibitem{GaltJM}
Nicolas Galtier and A.~Jean-Marie.
\newblock {M}arkov-modulated {M}arkov chains and the covarion process of of
  molecular evolution.
\newblock {\em J. Comput. Biol.}, 11(4):727--733, 2004.

\bibitem{GasGui}
Olivier Gascuel and St\'ephane Guindon.
\newblock Modelling the variability of evolutionary processes.
\newblock In Olivier Gascuel and Mike Steel, editors, {\em Reconstructing
  Evolution: New Mathematical and Computational Advances}, pages 65--107.
  Oxford University Press, 2007.

\bibitem{GRDH}
St\'ephane Guindon, Allen~G. Rodrigo, Kelly~A. Dyer, and John~P. Huelsenbeck.
\newblock Modeling the site-specific variation of selection patterns across
  lineages.
\newblock {\em P.N.A.S.}, 101:12957--12962, 2004.

\bibitem{hornJohnson85}
Roger~A. Horn and Charles~R. Johnson.
\newblock {\em Matrix Analysis}.
\newblock Cambridge University Press, 1985.

\bibitem{HuelCov}
John Huelsenbeck.
\newblock Testing a covariotide model of {DNA} substitution.
\newblock {\em Mol. Biol. Evol.}, 19:698--707, 2002.

\bibitem{MR1088949}
J.~B. Kruskal.
\newblock Rank, decomposition, and uniqueness for {$3$}-way and {$N$}-way
  arrays.
\newblock In {\em Multiway data analysis (Rome, 1988)}, pages 7--18.
  North-Holland, Amsterdam, 1989.

\bibitem{MR0488592}
Joseph~B. Kruskal.
\newblock More factors than subjects, tests and treatments: an indeterminacy
  theorem for canonical decomposition and individual differences scaling.
\newblock {\em Psychometrika}, 41(3):281--293, 1976.

\bibitem{MR0444690}
Joseph~B. Kruskal.
\newblock Three-way arrays: rank and uniqueness of trilinear decompositions,
  with application to arithmetic complexity and statistics.
\newblock {\em Linear Algebra and Appl.}, 18(2):95--138, 1977.

\bibitem{Petrie}
T.~Petrie.
\newblock Probabilistic functions of finite state {M}arkov chains.
\newblock {\em Ann. Math. Statist}, 40:97--115, 1969.

\bibitem{Rog01}
James~S. Rogers.
\newblock Maximum likelihood estimation of phylogenetic trees is consistent
  when substitution rates vary according to the invariable sites plus gamma
  distribution.
\newblock {\em Syst. Biol.}, 50(5):713--722, 2001.

\bibitem{MR2060009}
Charles Semple and Mike Steel.
\newblock {\em Phylogenetics}, volume~24 of {\em Oxford Lecture Series in
  Mathematics and its Applications}.
\newblock Oxford University Press, Oxford, 2003.

\bibitem{MR1604518}
Chris Tuffley and Mike Steel.
\newblock Modeling the covarion hypothesis of nucleotide substitution.
\newblock {\em Math. Biosci.}, 147(1):63--91, 1998.

\bibitem{Wang07}
Huai-Chun Wang, Matthew Spencer, Edward Susko, and Andrew Roger.
\newblock Testing for covarion-like evolution in protein sequences.
\newblock {\em Mol. Biol. Evol.}, 24(1):294--305, 2007.

\bibitem{Whelan}
Simon Whelan.
\newblock Spatial and temporal heterogeneity in nucleotide evolution.
\newblock {\em preprint, (2008)}.

\end{thebibliography}

\vfill
\end{document}